\documentclass[pre,aps,showpacs,twocolumn]{revtex4}
\usepackage{epsfig}

\newcommand{\be}{\begin{equation}}
\newcommand{\ee}{\end{equation}}
\newcommand{\bc}{\begin{center}}
\newcommand{\ec}{\end{center}}
\newcommand{\bi}{\begin{itemize}}
\newcommand{\ei}{\end{itemize}}
\newcommand{\ba}{\begin{eqnarray}}
\newcommand{\ea}{\end{eqnarray}}

\newcommand{\ignore}[1]{}

\begin{document}

\title{Nonequilibrium transitions in complex networks: a model of
social interaction}
\author{Konstantin Klemm}
\email{klemm@nbi.dk}
\author{V\'{\i}ctor M. Egu\'{\i}luz}
\homepage{http://www.imedea.uib.es/~victor}
\email{victor@imedea.uib.es}
\author{Ra\'ul Toral}
\email{raul@imedea.uib.es}
\author{Maxi San Miguel}
\email{maxi@imedea.uib.es}
\affiliation{Instituto Mediterr\'aneo de
Estudios Avanzados IMEDEA (CSIC-UIB), E07071 Palma de Mallorca
(Spain)}

\date{\today}

\begin{abstract}

We analyze the non-equilibrium order-disorder transition of
Axelrod's model of social interaction in several complex networks.
In a small world network, we find a transition between an ordered
homogeneous state and a disordered state. The transition point is
shifted by the degree of {\em spatial} disorder of the underlying
network, the network disorder favoring ordered configurations. In
random scale-free networks the transition is only observed for
finite size systems, showing system size scaling, while in the
thermodynamic limit only ordered configurations are always
obtained. Thus in the thermodynamic limit the transition
disappears. However, in structured scale-free networks, the phase
transition between an ordered and a disordered phase is restored.

\end{abstract}
\maketitle

%%%%%%%%%%%%%%%%%%%%%%%%%%%%%%%%%%%%%%%%%%%%%%%%%%%%%%%%%%%%%%%%%%%%%%%
\section{Introduction}

Lattice models are a powerful basic instrument in the study of phase
transitions in equilibrium Statistical Mechanics, as well as in non-equilibrium
systems \cite{Marro98}. Traditionally, equilibrium phase transitions have been
studied in regular lattices, with the critical temperature being a
non-universal quantity that depends on the particular lattice under
consideration, while critical exponents and some amplitude ratios are universal
quantities depending only on spatial dimension and some symmetries of the order
parameter. The detailed structure of the regular network connections is, in
most cases, irrelevant in the sense of the renormalization group. However,
recent research in the structure and topology of complex networks
\cite{rmp,aip} has shown that social interactions and, more generally,
biological and technological networks, are far from being regular as well as
being also far from a random network or from a mean-field network linking
all-to-all. This has triggered the study of standard models of Statistical
Mechanics in these complex networks. In particular, recent studies of the Ising
model in the so--called {\em small-world} \cite{Watts98} and the {\em
scale-free} \cite{Barabasi99} networks have shown that the behavior of the
model differs from that observed in a regular network.

A small world network \cite{Watts98} is generated by rewiring with
a probability $p$ the links of a regular lattice by long-distance
random links. The presence of a small fraction of ``short cuts''
connecting otherwise distant points, drastically reduces the
average shortest distance between any pair of nodes in network,
keeping the clustering high. The small-world networks generated by
rewiring links have degree distributions with exponential tails.
In contrast, scale-free networks \cite{Barabasi99} are
characterized by a fat-tailed (power law) degree distribution.
They are usually modelled by growing networks and preferential
attachment of links.

The Ising model in small-world topologies shows a change of
behavior from the regular case to the mean field characteristics.
In Ref.~\cite{Barrat00} it is shown analytically that for a small
world lattice, obtained from rewiring with probability $p$ the
links of a 1-d ring lattice with $2k$ nearest neighbors
interactions, the crossover temperature to the mean field critical
behavior varies for $p<<1$ as $T_{co}(p)\propto  -k(k+1)/\ln(p)$
whereas the critical temperature scales as $T_c(p)\propto
-2k/\ln(p)$, so that a ferromagnetic ordered phase exists for any
finite value of $p$. The crossover to mean field behavior in small
world ring lattices has been further discussed in
Refs.~\cite{Gitterman00,Hong02}, whereas numerical results in 2-d
and 3-d lattices have been reported in \cite{Herrero02}.
Interestingly, if directed links are considered, not only the
critical temperature changes but the nature of the transition also
switches from second to first order \cite{Sanchez02}.

A much different behavior is observed in scale-free networks. This
can be related to the influence of the presence of so-called hubs,
{\em i.e.\ } units whose degree is much larger than average. This
is well illustrated by the behavior of the Ising model in
scale-free networks with degree distribution $P(k) \propto
k^{-\gamma}$, $\gamma>1$. The results of Refs.~\cite{Aleksiejuk02}
and \cite{Goltsev02} show that equilibrium systems exhibit a phase
diagram that is qualitatively different from the mean-field case.
In particular, the Ising model in a random scale-free networks
shows an infinite critical temperature in the thermodynamic limit
of an infinite number of nodes. In fact it has been developed
analytically a mean field theory connecting the exponent of the
degree distribution and the critical behavior of the Ising model
\cite{Dorogovtsev02,Leone02,Bianconi02}.

In this paper we address the question of the role played by the
topology of complex networks in non-equilibrium transitions of
models in which there is interaction between the variables
associated with the nodes connected by links in the network. This
is a natural next step beyond the analysis of equilibrium,
Ising--type models  in these complex networks. Simple
non-equilibrium models closely related to percolation have been
already considered \cite{Kuperman01,Pastor01,Eguiluz02}. Here, and
given the social motivation and relevance of these complex
networks, we have chosen to analyze the model proposed by Axelrod
for the dissemination of culture \cite{Axelrod97}. The spreading
process in this model cannot be reduced to a percolation process.
The model rather describes a competition between dominance and
spatial coexistence of different states in a non-equilibrium
dynamics of coupled Potts-type models. The model was originally
considered by Axelrod in a square lattice. The Statistical
Mechanics analysis of the model in this regular two-dimensional
network identifies a nonequilibrium order-disorder phase
transition \cite{Castellano00}. However, it is interesting to
notice that, in his original paper, Axelrod already discussed the
relevance of the topology, speculating that ``With random
long-distance interactions, the heterogeneity sustained by local
interaction cannot be sustained.'' In particular we consider here
this question.

In the next Section we introduce the original model proposed in
Ref.~\cite{Axelrod97} and summarize briefly the main results in
regular 2-d networks.  The model in small-world and scale-free
networks is presented in Section~\ref{s:SW} and \ref{s:SF}
respectively. The non-equilibrium transition is shown to disappear
in the thermodynamic limit of a scale free network. We then
consider in Section \ref{s:MSF} a structured scale free network
\cite{Klemm02} which shares characteristics of small-world and
scale-free networks. A non-equilibrium transition is shown to
persist for large systems in this network. Our conclusions are
summarized in Section~\ref{s:conclusions}.

%%%%%%%%%%%%%%%%%%%%%%%%%%%%%%%%%%%%%%%%%%%%%%%%%%%%%%%%%%%%%%%%%%%%%%%
\section{The model}

The model we study is defined \cite{Axelrod97} by considering $N$
agents as the sites of a network. The state of agent $i$ is a
vector of $F$ components (cultural features)
$(\sigma_{i1},\sigma_{i2},\cdots,\sigma_{iF})$. Each $\sigma_{if}$
can take any of the $q$ integer values (cultural traits) $1,\dots,q$,
initially assigned independently and with equal probability $1/q$.
The time-discrete dynamics is defined as iterating the following
steps:
\begin{enumerate}
\item Select at random a pair of sites of the network connected by a bond $(i,j)$.
\item Calculate the {\em overlap} (number of shared features)
$l(i,j) = \sum_{f=1}^F \delta_{\sigma_{if},\sigma_{jf}}$.
\item If $0<l(i,j)<F$, the bond is said to be {\sl active} and sites
$i$ and $j$ interact with probability $l(i,j)/F$. In case of interaction,
choose $g$ randomly such that $\sigma_{ig}\neq\sigma_{jg}$ and set $\sigma_{ig}=\sigma_{jg}$.
\end{enumerate}

In any finite network the dynamics settles into an {\em absorbing}
state, characterized by the absence of active bonds. Obviously all
the $q^F$ completely homogeneous configurations are absorbing.
Homogeneous means here that all the sites have the same value of
the cultural trait for each cultural feature. Inhomogeneous states
consisting of two or more homogeneous domains interconnected by
bonds with zero overlap are absorbing as well. A domain is a set
of contiguous sites with identical cultural traits. We define an
order parameter in this system \cite{Castellano00,Klemm02b} as the
relative size of the largest homogeneous domain $S_{max}/N$, being
$N$ the number of sites in the network.

Previous results have been obtained in square lattices with
nearest neighbor interaction. A variation of the model with
initial distribution of traits according to a Poisson rather than
a uniform distribution shows a non-equilibrium order-disorder
phase transition where the number of traits $q$ plays the role of
a control parameter \cite{Castellano00}. The system reaches
ordered absorbing states for $q<q_c$ ($S_{max}={\cal O}(N)$) and
disordered states for $q>q_c$ ($S_{max}\ll N$). The same type of
phase transition occurs in the original model with a uniform
initial distribution of traits \cite{Klemm02b}.

%%%%%%%%%%%%%%%%%%%%%%%%%%%%%%%%%%%%%%%%%%%%%%%%%%%%%%%%%%%%%%%Fig 1
\begin{figure}
\centerline{\epsfig{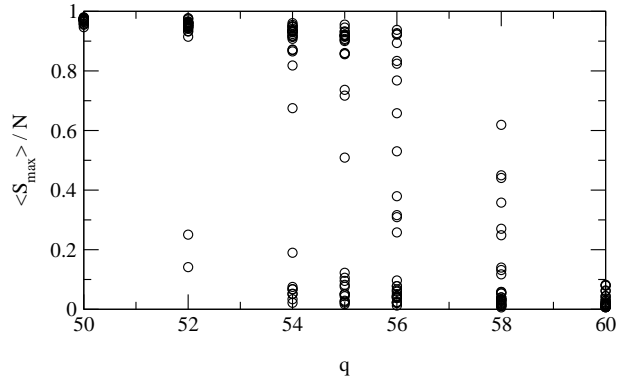}}
\caption{\label{realizations_reg} The order parameter $
S_{max}/N$ as a function of $q$ in regular lattices with
$N=100^2$ nodes for $F=10$. For each value of $q$ the outcome of
32 independent realizations is shown. The transition occurs for $q
\approx 55$ (see Fig.\ref{hist_reg}). }
\end{figure}

\begin{figure}
\centerline{\epsfig{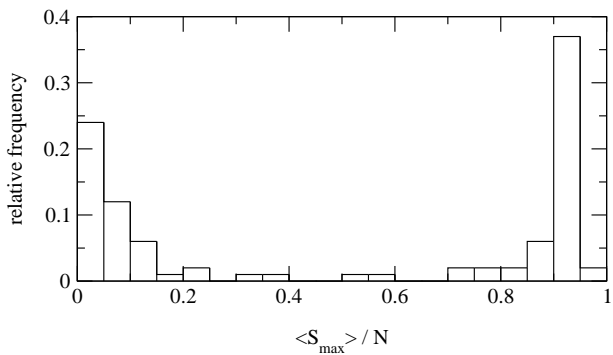}}
\caption{\label{hist_reg}
Distribution of the order parameter at $q=55$, $F=10$ in a square
lattice of size $N=100^2$. The distribution is based on 100 independent
realizations.
}
\end{figure}

When comparing the effect of complex networks in this phase
transition with the equilibrium Ising transition one should notice
several conceptual differences. First, this is a sort of
zero-temperature transition in which ordered or disordered states
exists with no reference to thermal fluctuations. In fact, the
effect of small noise in this system is essential, revealing the
presence of metastable states and changing the phase diagram in a
nontrivial way \cite{Klemm02b}. A second related point is that the
control parameter of the transition $q$, is here not a collective
property of the system as temperature, but rather an ingredient of
the definition of the system itself. In a way, the transition
occurs going from one system to another as $q$ is changed.
Finally, and in reference to critical properties and exponents, we
note that the transition (except for F=2) is of first-order type:
In figure \ref{realizations_reg} we plot the final values for the
order parameter, obtained for $32$ different realizations of the
dynamics. Notice that for $q<50$ all the systems end up in a
homogenous state that basically fills up the entire lattice
($S_{max}/N\approx 1$) whereas for $q>60$ the maximum homogenous
regions obtained are very small. This is the order--disorder phase
transition discussed before. For $50<q<60$ we observe bistability
in the sense that the system settles around any of two mean values
for the order parameter. This bistability, which is usually
associated with first order phase transitions, is clearly made
explicit in the corresponding histogram shown in figure
\ref{hist_reg} where the two preferred values appear as maxima of
the histogram. The transition point corresponds to $q=q_c$ for
which these two values are equally probable.

%%%%%%%%%%%%%%%%%%%%%%%%%%%%%%%%%%%%%%%%%%%%%%%%%%%%%%%%%%%%%%%%%%%%%%%
\section{Small world network}
\label{s:SW}

Social networks are far from being regular or completely random.
However they also share some features with them. On the one hand,
social networks are known to be {\em small} \cite{Milgram67}, i.e.
any pair of nodes in the network can be connected following a
number of links much smaller than the size of the network. This is
also observed in random networks, where the average shortest
distance between pair of nodes (the so called path length $\ell$)
increases logarithmically with the size of the network $\ell \sim
\ln N$, while in regular lattice in $d$-dimensions $\ell \sim
N^{1/d}$. On the other hand, social networks are also known to
form cliques \cite{Wasserman94}, i.e. groups of nodes highly
connected between them. ``Cliquishness'' can be characterized by
the so-called clustering coefficient $C$, which is defined as the
relative number of closed triangles in the network. Regular
lattices can show large clustering while in random networks $C
\sim N^{-1}$. High clustering and short path length define a small
world network.

% SW %%%%%%%%%%%%%%%%%%%%%%%%%%%%%%%%%%%%%%%%%%%%%%%%%%%%%%%%%%%%%%%%%%%%%%
\begin{figure}
\centerline{\epsfig{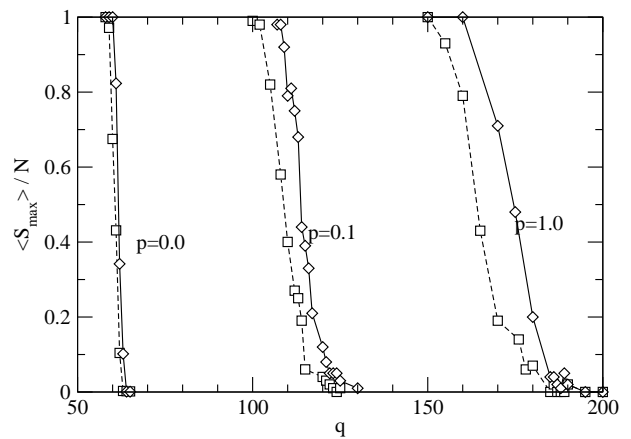}}
\caption{\label{fsw1} The average order parameter $\langle
S_{max}\rangle/N$ as a function of $q$ for three different values
of the small world parameter $p$. System sizes are $N=500^2$
(squares) and $N=1000^2$ (diamonds), number of features $F=10$.
Each plotted value is an average over $100$ runs with independent
rewiring ($p>0$) and independent initial conditions. }
\end{figure}

The first model encompassing the small world effect was introduced
in Ref.~\cite{Watts98} proposing an algorithm that interpolates
between a random and a regular lattice. First one generates a
two-dimensional regular lattice with bonds between nearest
neighbors and open boundary conditions. Then for each bond $(ij)$,
with probability $p$ detach the bond from node $j$ and attach it
to a node $l$ instead. Node $l$ is chosen at random with the
restriction that duplicate and self-connections are excluded. The
parameter $p$ interpolates between the original regular lattice
($p=0$, no rewiring) and a network very similar to a random graph
($p=1$). Thus in the limiting case $p=0$ we have a network with
high clustering but also large path length; in the limit $p=1$ we
have networks with small path length but also small clustering.
For intermediate values of $p$ the algorithm generates networks
with high clustering and small path length.

% SW %%%%%%%%%%%%%%%%%%%%%%%%%%%%%%%%%%%%%%%%%%%%%%%%%%%%%%%%%%%%%%%%%%%%%%
\begin{figure}
\centerline{\epsfig{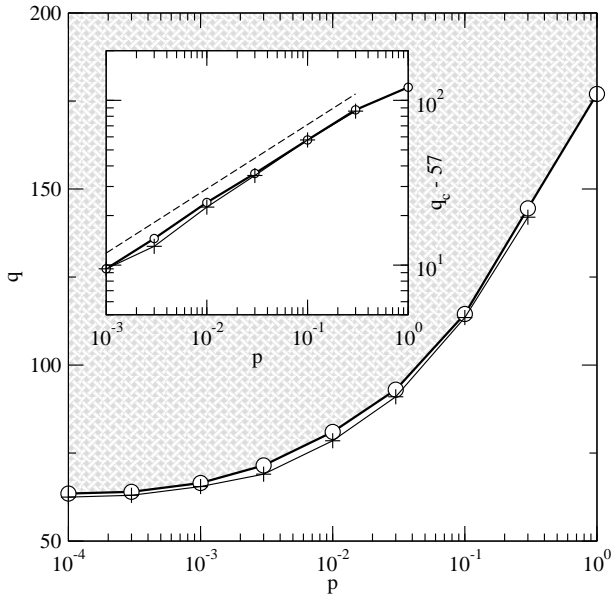}}
\vskip 1.cm
\caption{\label{fsw2} Phase diagram for the Axelrod
model in a small world network. The curve separates parameter
values (p,q) which produce a disordered state (shaded area) from
those with ordered outcome (white area). For a given $p$ the
plotted value $q_c$ is the one for which the value of the order
parameter is closest to the, somewhat arbitrary but small, value
$0.1$ for system size $N=500^2$ and $F=10$. Inset: After
subtraction of a bias $q_c (p=0)=57$, $q_c(p)$ follows a power law
$\propto p^{0.39}$ (dashed line). }
\end{figure}

We now study the behavior of Axelrod's model in dependence of $p$.
A small world network is used from the beginning of each
simulation run. Figure \ref{fsw1} shows the dependence of the
order parameter on $q$, for three different values of $p$. For any
fixed value $p>0$ we find a nonequilibrium phase transition which
becomes sharp and well defined as the system size increases. There
is a critical value $q_c$ of the control parameter which separates
the ordered and the disordered state, just as in regular lattices.
However, $q_c$ increases with the amount of spatial disorder. This
is clearly shown in the $(p,q)$-phase diagram, Fig. \ref{fsw2}.
The filled area above the $(p,q_c(p))$-curve represents the
disordered states, the area below the curve represents the ordered
states. Consequently, for values $q<q_c(p)$ the outcome of the
dynamics is always complete order, whereas for $q>q_c(p)$ only
disordered frozen states are encountered. The density $p$ of
rewired bonds determines the nature of these frozen states, but
for $q<q_c(p=1)$ the system orders by increasing $p$, that is, the
number of long distance links. We find a dependence
$q_c(p)-q_c(p=0) \propto p^\alpha$ with $\alpha=0.4$ obtained from
a best fit. This result is displayed in the inset of Fig.\
\ref{fsw2}. Therefore, we find the same qualitative result that
for the equilibrium Ising model, in the sense that the small world
connectivity favors ordered states.

The robustness of the phase diagram is shown by performing a
different dynamical scenario. First, a run of the dynamics in a
regular lattice is performed. Only after an absorbing
configuration has been reached the lattice is rewired according to
the above rewiring procedure with the parameter $p$. After the
rewiring, the configuration is not necessarily frozen because the
rewiring can introduce active bonds connecting compatible cultures
that have been disconnected before. Starting the dynamics again,
the system may relax to a different absorbing configuration, which
in general is more ordered than the configuration reached before
the rewiring. After this second phase of relaxation, the order
parameter is measured in the absorbing state. We find that the
results of this alternative scenario (see Fig.\ \ref{fsw2}) are in
good agreement with the ones of the above original scenario,
starting with a small world network in the initial condition.

%%%%%%%%%%%%%%%%%%%%%%%%%%%%%%%%%%%%%%%%%%%%%%%%%%%%%%%%%%%%%%%%%%%%%%%
\section{Scale-free networks}
\label{s:SF}

One important ingredient missing in the small world networks considered
so far is that the degree distribution does not show a fat tail.
Although it is not clear whether social networks present a power law
distribution of degree, the evidence indicates that they are ubiquitous
in biological and artificial networks \cite{Amaral00,Albert02}.
Scale-free networks are characterized by a power law tail in the degree
distribution of the form $P(k) \propto k^{-\gamma}$ where the exponent
$\gamma$ lies in the range between 2 and 3. Two ingredients have been
shown to be sufficient to generate such feature: growing number of
nodes and preferential attachment of links.

% SF %%%%%%%%%%%%%%%%%%%%%%%%%%%%%%%%%%%%%%%%%%%%%%%%%%%%%%%%%%%%%%%%%%%%%%
\begin{figure}
\centerline{\epsfig{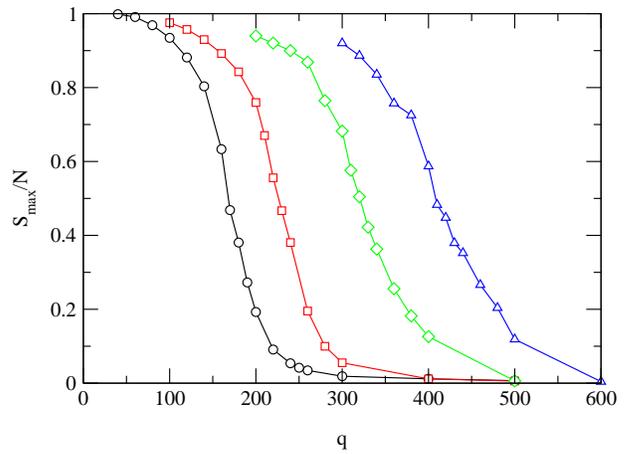}} \caption{\label{fsf1} The average order parameter
$\langle S_{max}\rangle/N$ in random scale-free networks for
$F=10$. Averages are taken over 1000 independent realizations.
Different curves are for different system sizes: 1000 (circles),
2000 (squares), 5000 (diamonds), and 10000 (triangles).}
\end{figure}

\begin{figure}
\centerline{\epsfig{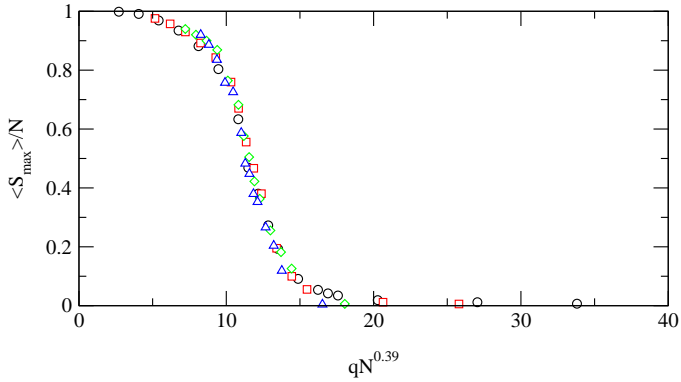}} \caption{\label{fsf1_2} Rescaled plot of the data shown
in Fig. \ref{fsf1} for different system sizes.}
\end{figure}

\begin{figure}
\vspace{1cm}
\centerline{\epsfig{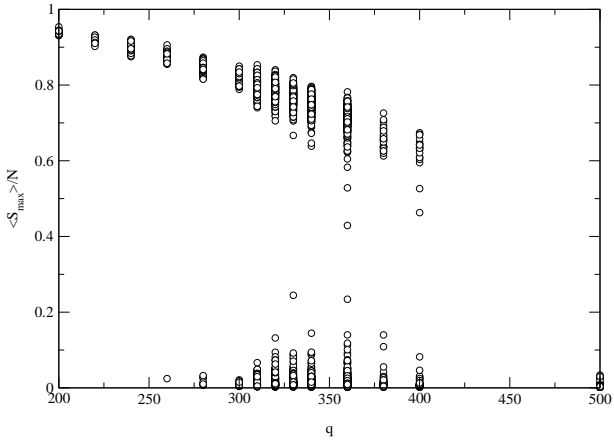}}
\caption{\label{fsf2} Order parameter in random scale-free networks of
size $N=5000$. For each value of $q$ ($F=10$) the outcomes of 100
independent realizations are shown.
}
\end{figure}

The well established Barab\'asi--Albert model is based in these
two mechanisms \cite{Barabasi99}. At each time step a new node is
added to the network and attaches $m$ links to an existing node
with degree $k$ with probability $\Pi (k) \propto k$. This
algorithm generates networks whose degree distribution follows
$P(k) = 2 m^2 k^{-3}$, the path length increases as $\ell \sim \ln
N$, and the clustering decreases as $C \sim (\ln N )^2 /N$. We
have studied the dynamics of Axelrod's model for the diffusion of
culture in scale-free networks following this algorithm. In
Fig.~\ref{fsf1} we show our results for the order parameter for
different system sizes. For a given size $N$ we find a transition
at $q_c(N)$. We can define the critical value $q_c(N)$ as the
value where the standard deviation of the distribution of $S_{\rm
max}/N$ reaches the maximum value. We find that $q_c(N) \sim
N^{0.39}$. Using this result we observe data collapse with a
rescaling $q N^{-\beta}$, see Fig.~\ref{fsf1_2}. The best result
is obtained for $\beta=0.39$ in excellent agreement with the
scaling obtained previously. This indicates that in the
thermodynamic limit the transition disappears and the ordered
monocultural state establishes in the system. This behavior is
similar to the Ising model in regular and scale-free networks:
While in a two-dimensional lattice the Ising model displays a
phase transition at a finite critical temperature, in random
scale-free networks an effective transition is observed for finite
systems where the effective critical temperature diverges
logarithmically with system size. This can be explained by the
role of the hubs (nodes with a large number of links) in these
networks. They are highly instrumental in establishing
ferromagnetic order in the system. The same prominent role is
played by the hubs in the case of the dissemination of culture.
The hubs help the spreading of cultural traits as can be inferred
from the observed dependence with system size. Note, however that
the effective transition of Axelrod´s model for a finite system in
a scale free network displays the characteristics of a first order
transition: We show in Figs. \ref{fsf1} and \ref{fsf2} the same
type of behavior observed in Figs. \ref{realizations_reg} and
\ref{hist_reg} for the regular network. For a range of values of
$q$ around $q_c$ a realization ends either in an ordered
monocultural state or in a disordered frozen configuration, the
two preferred values of the order parameter clearly seen in the
histogram of Fig. \ref{hist_reg}.

% SF %%%%%%%%%%%%%%%%%%%%%%%%%%%%%%%%%%%%%%%%%%%%%%%%%%%%%%%%%%%%%%%%%%%%%%

%%%%%%%%%%%%%%%%%%%%%%%%%%%%%%%%%%%%%%%%%%%%%%%%%%%%%%%%%%%%%%%%%%%%%%%

%%%%%%%%%%%%%%%%%%%%%%%%%%%%%%%%%%%%%%%%%%%%%%%%%%%%%%%%%%%%%%%%%%%%%%%
\begin{figure}
\centerline{\epsfig{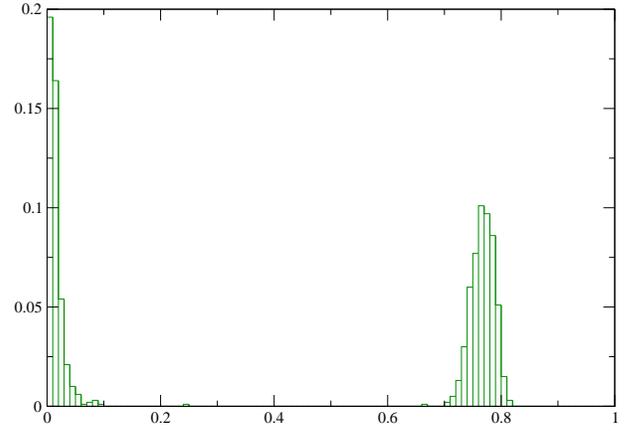}}
\caption{\label{fsf3} Distribution of the order parameter in scale-free
networks of size $N=5000$ for $q=360$, and 1000 realizations.
}
\end{figure}

\section{Structured scale-free networks}
\label{s:MSF}

The scale-free networks considered in the previous section underestimate
the clustering observed in real networks \cite{Ravasz02}. A question
that merits to be addressed is if this low clustering coefficient is
responsible for the absence of the phase transition in the
thermodynamic limit. In order to reproduce a high clustering along with
a scale-free distribution of the degree, we employ the
networks generated by the algorithm proposed in Ref.~\cite{Klemm02}:
Again at each time step we add a new node to the
network. The node is attached to the $m$ active nodes in the network.
The new node becomes active and one of the $m+1$ active nodes is
deactivated with probability $\Pi (k) \propto k^{-1}$. Starting from
$m$ fully inter-connected active nodes, this algorithm generates scale-free
networks with a clustering coefficient $C \approx 5/6$ independent of
system size. In the following we call these networks {\em structured}
scale-free networks because of the large clustering coefficient, the
strong negative correlation between degrees of connected nodes
\cite{Eguiluz02}, and the modular structure \cite{Ravasz02}. These
properties are not found in the {\em random} scale-free networks of
the previous section.

% HCSF %%%%%%%%%%%%%%%%%%%%%%%%%%%%%%%%%%%%%%%%%%%%%%%%%%%%%%%%%%%%%%%%%%%%%%
\begin{figure}
\centerline{\epsfig{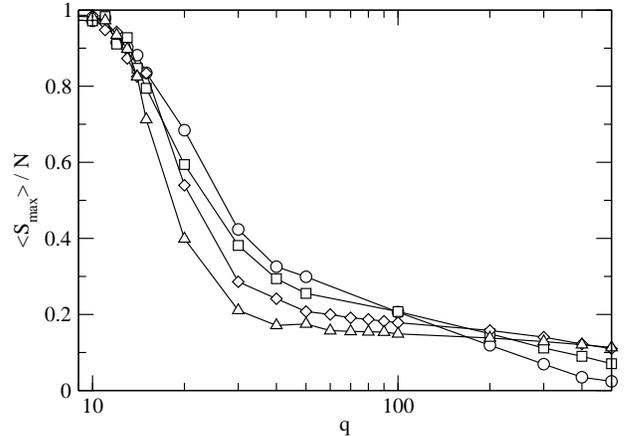}} \caption{\label{fhcsf0} The average order parameter
$\langle S_{max}\rangle/N$ as a function of $q$ for $F=10$ in
structured scale-free networks. The networks contained $N=1000$
(circles), $N=2000$ (squares), $N=5000$ (diamonds), and $N=10000$
(triangles) nodes with $F=10$ features. Each data point is an
average over 32 independent realizations. }
\end{figure}
%%%%%%%%%%%%%%%%%%%%%%%%%%%%%%%%%%%%%%%%%%%%%%%%%%%%%%%%%%%%%%%%%%%%%%%
\begin{figure}
\centerline{\epsfig{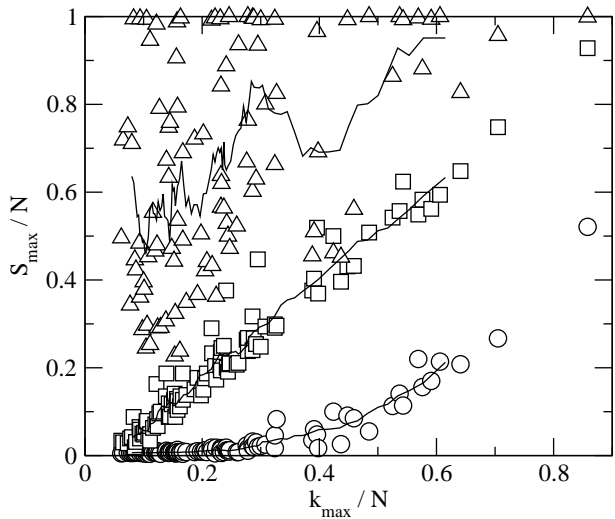}}
\caption{\label{fhcsf1}
Relation between the size of the largest cultural cluster and the largest
degree in the network for $q=20$ (triangles), $q=100$ (squares) and $q=500$
(circles). Each data point is the outcome of one realization run in a
structured scale-free network of size $N=1000$. Solid lines are running
overages over 10 adjacent data points for $q=20,100,500$ (top to bottom).
For each value of $q$, 100 independent networks and initial conditions
were generated.
}
\end{figure}

As shown in Fig.~\ref{fhcsf0}, in the structured scale-free networks
the model displays a behavior different to what we observed in random
scale-free networks in the previous section. For $q \lesssim 10$ the
system settles into an ordered state. For increasing values of $q$, the
order parameter undergoes a decay whose slope grows with system size.
This suggests a phase transition at $q_c\approx 10$, in contrast to the
absence of a transition point found for the randomly wired scale-free
networks in the thermodynamic limit. As on large scales the structured
scale-free networks have one-dimensional topology \cite{Klemm02a} it
seems natural that this transition at $q_c \approx F$ coincides with the
behavior of the model found in one-dimensional regular lattices
\cite{Klemm02c}.

At difference with the regular lattices, in the structured scale-free
networks for $q > q_c$ the order parameter does not tend to zero. It
reaches a finite plateau value, indicating partial ordering of the
system. Only for values $ q \gg q_c$ the order parameter drops below
the plateau value and tends to zero. This behavior may be understood by
relating the size $S_{max}$ of the largest cultural cluster with the
largest degree $k_{max}$ present in the network, as shown in
Fig.~\ref{fhcsf1}. In the intermediate range $050<q<200$, where the
plateau of the order parameter is observed, we find $S_{max} \approx
k_{max}$ for almost all realizations. This suggests that the largest
hub, the node with the largest degree, and its neighbors order such
that they form the largest cluster in the absorbing state. As $q$ is
reduced and its value approaches $q_c$ from above, the ordering goes
beyond the largest hub and an increasing part of the network forms the
largest cluster. On the other hand, for large values $q>200$, the
neighborhood of the largest hub does no longer reach complete ordering
and $S_{max} < k_{max}$. The $q$ value for the onset of the decay of
$S_{max}$ below $k_{max}$ is expected to be dependent on system size:
as Fig.~\ref{fhcsf0} shows, for increasing system size, the plateau in
the order parameter extends to larger values of $q$. With inrceasing
system size, the value of the plateau is expected to decrease as
$k_{max}/N = N^{\beta-1}$ with $\beta = (\gamma-1)^{-1}$, where $\gamma$
is the exponent of the degree distribution. These results suggest that
in the limit $N \to \infty$, the dynamics of the social interaction model in
structured networks experiences a transition similar the one observed in
a one-dimensional lattice.

%%%%%%%%%%%%%%%%%%%%%%%%%%%%%%%%%%%%%%%%%%%%%%%%%%%%%%%%%%%%%%%%%%%%%%%
\section{Conclusions}
\label{s:conclusions}

We have found that the nonequilibrium transition between order and
disorder that exists in a regular d=2 network for Axelrod´s model
of cultural influence \cite{Axelrod97} is modified by underlying
complex networks with similar qualitative features that an
equilibrium thermal Ising-type transition. We have shown that the
transition pertains also in the presence of random long-distance
connections: with increasing density of long-distance connections
in a small world network, the critical point $q_c(p)$ increases.
Therefore the small world connectivity favors cultural
globalization as described by the ordered state. The value of
$q_c$ reaches a maximum for the random network obtained with a
$p=1$ probability of rewiring in the small word network
construction. A transition from disorder to order is obtained
increasing $p$ for a fixed value of the control parameter
$q<q_c(p=1)$. We have also found that, for a fixed finite system
size, the scale free connectivity is more efficient than the
limiting random connectivity of the small world network in
promoting the ordered state of cultural globalization. However,
there is a system size scaling in the transition observed for a
free scale network, so that the transition disappears in the
thermodynamic limit: In the presence of scale-free interactions
the order state prevails due to the presence of hubs. The
consideration of structured scale free-networks restores the
order-disorder transitions in spite of the hubs, but the value of
the order parameter for the disordered state reveals the existence
of ordered clusters.

%%%%%%%%%%%%%%%%%%%%%%%%%%%%%%%%%%%%%%%%%%%%%%%%%%%%%%%%%%%%%%%%%%%%%%

\end{document}